\documentclass[12pt,longbibliography]{article}
\usepackage{verbatim}
\usepackage{lineno}
\usepackage{amsmath}
\date{}

\usepackage[hyperindex=true,citecolor=black,breaklinks=true,colorlinks=true,linkcolor=black]{hyperref} %controla as cores das ligazóns de número de ecuacións, etc.
\usepackage{mathtools}
\usepackage{multirow}
\usepackage{braket}
\usepackage{caption}
\usepackage{subcaption}

\title{Multichannel Hybrid Quantum Cryptography for Submarine Optical Communications}

\author{Jesús Liñares$^{1}$, Xesús Prieto-Blanco$^{1}$,  \\ Alexandre Vázquez-Martínez$^{1}$  and  Eduardo F. Mateo$^{2}$
\\
 \small (1)\,Optics Area.  Applied Physics Department.  QMatterPhotonics Research Group \\ \small and iMATUS Institute.    Campus Vida. Universidade de Santiago de Compostela,  \\ \small E-15782 Santiago de Compostela, Galicia, Spain\\ 
  \small  (2)\,Nippon Electric Company (NEC). Minato-ku. Tokyo. Japan\\
}

\begin{document}
%is required; see note below about the corresponding author designation
\maketitle
% use {asbstract*} to suppress the copyright line. Copyright information will be added in production
\begin{abstract} 
%100 palabras
We present a multichannel hybrid  quantum cryptography approach intended for submarine quantum optical communications between  Alice and Bob separated by a distance beyond the current QKD possibilities, each located on a coastline. It is based on the difficulty of a  simultaneous  access to $M$ optical submarine  channels. The optical lines from the coastline and ideally to the end of   the continental shelf  are  governed by the quantum  properties of the light under  an autocompensating high-dimensional discrete-modulation continuous variable QKD  protocol. The hybrid approach consists of combining  several secret keys of  the  $M$ channels   and  introducing  extra layers of security, passive and/or active,  on the  non quantum optical lines located beyond the continental shelf.  

\end{abstract}

%%%%%%%%%%%%%%%%%%%%%%%%%%  body  %%%%%%%%%%%%%%%%%%%%%%%%%%
\section{Introduction}

The development of quantum optical communications is a major challenge before the imminence of quantum supremacy. Indeed, the security and confidentiality of communications is nowadays implemented by encryption methods that rely on mathematical problems that are computationally prohibitive to decrypt. However, these problems (such as factorization or large prime numbers) can be easily solved by quantum computers. This threat to security can be solved by utilizing Quantum Key Distribution (QKD)  \cite{Pirandola2020}. QKD lines uses unique statistical features of quantum states which reveal, unconditionally, the presence of an eavesdropper. Usually, these lines are terrestrial ones with optical fibres or free space lines which have a determined level of security and secret key rates which decrease abruptly from a determined transmission distance. \textcolor{black}{Secure distances of a few hundred kilometres can be achieved, but high key rates in continuous variable protocols are rather limited to distances of up to several tens of kilometres \cite{Adnan2024_CV-QKD_100km,Bian2025_40-km_Mbps_DM-CV-QKD}.} Practical optical networks require many access points to monitor the performance and integrity of the transmission links. QKD provides the means to stablish secure communications in such fibre plants, where an eavesdropper can easily access the data. One of the limitations of QKD is the incompatibility with optical amplifiers, which limits the transmission length of the key \cite{Bai2012}. For that, complex topologies of secure-nodes are proposed to regenerate the cryptographic keys \cite{Salvail2010}. Although such secure-nodes do not provide unconditional security and require access ports for maintenance and monitoring, they are an acceptable approach for long-distance QKD in practical optical networks while new technologies  such the quantum repeaters are not available.

On the other hand, submarine networks deliver 98\% of transcontinental traffic \textcolor{black}{through optical lines thousands km long \cite{chesnoy2025undersea}. Each modern cable contains between 8 and 48 single-mode fibre-pairs, depending on the application, and one of them can be allocated for key distribution. Commercial few-core optical fibre cables have already been installed, but their use is not yet widespread. Submarine optical cables} are, in nature, very well protected against eavesdropping with respect to terrestrial or satellite communications. First, both fibre and amplifiers are inaccessible from the outside as they are tightly encapsulated in cables and housings supporting water depths up to 8000m. Additionally, submarine networks are heavily electrified since they are powered from the landing stations, often withstanding voltages up to 18kV. \textcolor{black}{They supply sealed repeaters equipped with EDFAs (erbium-doped fibre amplifiers), installed at intervals of 50--100\,km. 
The first/last repeaters are typically a few tens of km from the shore. The protection of the line is different between shallow and deep waters whose demarcation is typically 1000\,m to 1500\,m water-depth.
}
In deep waters, submarine systems are protected by kilometres of water pressure whereas in shallow waters, submarine cables and repeaters are typically armored with thick steel sections and buried 3--10\,m deep in the seafloor. % in paths where the water-depth is below 1000m. 
Inside the cable, fibre-independence is guaranteed by the submarine system supplier. On top of this natural protection, sensing techniques such as DAS (Distributed Acoustic Sensing) are being proposed to detect vibrations and potential damages or attacks to the cable \cite{Chen2022}, and other types  physical layer security  methods could be used to maximize protection \cite{Alda2020}. \textcolor{black}{
In general, the sections of the optical line that are most accessible ---and therefore most vulnerable--- are those approaching the coastline, particularly those lying over the continental shelf, which is less than 100 m in deep. Its average width is about 80 km from the coast, although this value varies considerably. In certain regions, the shelf may reach several hundred kilometres offshore, whereas in others it is virtually absent.}

In this work, we propose a hybrid approach \textcolor{black}{for distributing cryptographic keys between distant nodes connected through existing submarine networks.
The key is derived from a combination of keys transmitted in parallel over several optical lines, which may be separated by many kilometres while interconnecting these nodes.
Each line includes a central section carrying classical signals, flanked on both ends by two quantum optical segments located on the submarine continental shelf. Depending on the bathymetry, in some areas the quantum segment may extend to the continental rise or even the abyssal plain. When this is not feasible, quantum sections can be followed by classical ones reinforced with additional physical security layers designed to detect or hinder any intrusion. This extra protection may cover a greater distance, for instance, reaching into deep waters.}
%several optical classical lines  which will receive combined secret keys from the quantum lines providing an extra physical layer of security.  As commented, such optical lines are  protected in a natural way (water pressure, buried lines, tight encapsulation\ldots) and we  also propose a  protection with additional physical layers  of security over  a certain distance before reaching deep waters (e.g., before a  depth of 1000\,m).  
Therefore, such approach  requires that Eve implements  simultaneous and complex attacks  to several optical  lines in deep waters and separated a certain distance.  Moreover, for the quantum lines we propose a discrete-modulation continuous variable QKD (DM-CV-QKD) protocol  with product weak coherent states (multimode states) measured by homodyne detection, which is usual in optical communications. The protocol is based on the seminal work by Namiki and Hirano for a single-mode weak coherent state \cite{Namiki2003}, therefore it will correspond to a high dimensional protocol  which also provides a larger quantum security \cite{Dynes2016}. Moreover, the QKD protocol will be a plug and play one, that is, an autocompensating protocol \cite{Bal19}, and therefore we get the advantage to locate both the quantum sources and detectors at terrestrial points.

The plan of the paper is as follows. In section 2 we present, by taking into account the general topology of a submarine optical communication line, an autocompensating DM-CV-QKD  protocol for two modes on the continental shelf; two modes is enough to show the security properties of the high-dimensional protocol and  its generalization to $N$ modes (multimode case) is straightforward. Moreover,  we show that the protocol  can also be used between points located  at both sides of the sea if the part of the line located in deep waters uses classical signals (hybrid protocol).  Next, in section 3, we present a hybrid approach for QKD with $M$ submarine channels (multichannel approach), that is, quantum channels connected to  submarine non quantum channels located at deep waters with  extra  physical layers of security based on combining  secret keys. In section 4 some examples of additional physical layers of security (APLS) will be presented for the submarine non-quantum channels before reaching deep waters. In section 5 conclusions are presented.

\section{Autocompensating HD-DM-CV-QKD protocol}
\begin{figure}[htbp]
\centering\includegraphics[width=11.5cm]{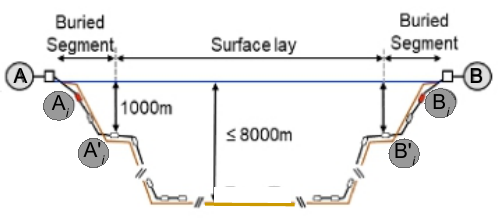}
\caption{Sketch of a submarine optical communication line. It is shown A and B, and arbitrary points with quantum lines A-A$_{i}$ and B$_{i}$-B. Moreover, points  $A'_{i}$ and $B'_{i}$ are considered to implement possible additional physical layers of  security in optical lines A$_{i}$-A'$_{i}$ and/or  B'$_{i}$-A$_{i}$.}
\label{Figinicial}
\end{figure}

First of all, we show in Fig.\ref{Figinicial} a sketch of a submarine optical communication line. There are  two terrestrial points A  and B to share a key. We assume that A (B) can be  connected by means of  quantum lines to any subsystem  A$_{i}$ (B$_{i}$) located at the continental shelf, \textcolor{black}{with A--A${i}$ (B--B${i}$) distances of approximately 30--40 km}. Next, a non quantum optical line starts, which has a buried segment of  the optical fibre line, and then the line continues on surface lay.  
As it will be depicted, although the non-quantum buried segment will be protected with the hybrid approach, it can also be reinforced with additional physical layers of security; for this purpose, supplementary points A'$_{i}$ and B'$_{i}$ are also indicated in Fig.\ref{Figinicial}, \textcolor{black}{with A$_{i}$--A'$_{i}$ and B'$_{i}$--B$_{i}$ distances of up to 100 km}. Finally, the lines A$'_{i}$-B'$_{i}$ will be protected, in principle, with the multichannel hybrid approach, as studied in Section 3, and with the own deep waters. \textcolor{black}{Note that through the mentioned non quantum lines only classical signals will be sent.} 

\textcolor{black}{In this section we present an autocompensating HD-DM-CV-QKD  protocol to be used  at each system  A-A$_{i}$ and B$_{i}$-B  located on the continental shelf along with the study of both  their QBER and Security Key Rate (SKR);  A$_{i}$ and B$_{i}$ will be connected by classical signals (hybrid scheme). We must stress that, as it will be explained in Section 3, an analogous  hybrid scheme is used in Earth-Satellite QKD systems \cite{Bedington2017, Liao2018}. The autocompesation method  is introduced to reduce  the technological complexity  under the sea. Additionally, we present an alternative  hybrid protocol for the entire subsea line between A and B, that reduces once again the technological complexity, however, as shown in subsection 2.4,  it also reduces in an effective way the SKR by a factor 1/2 with respect to the protocol used at  each quantum line A-A$_{i}$  or B$_{i}$-B on the continental shelf.}

\subsection{High-Dimensional DM-CV-QKD protocol on the continental shelves}\label{2-1}
The first practical aspect is worth discussing about our method, and  is common to any QKD method, corresponds  to maximizing the transmission distance and key-rate product, what we will made with a High-Dimensional (HD) DM-CV-QKD protocol which is fully compatible with the current optical communication technology. It is important to stress  that HD QKD protocols increase the secure key rate, and usually they are based on quantum superposition states in discrete variable \cite{Sheridan2010}, however in this case we will use product states which are much more compatible with continuous variable. We will explain the fundamentals of this HD-DM-CV-QKD protocol between two general points $\mathcal{A}$ and $\mathcal{B}$,  \textcolor{black}{in particular  between  points at the continental shelf A and A$_{i}$, or B and B$_{i}$}. For sake of expository convenience, we only consider two spatial modes, although it can easily be generalized to $N$ modes (HD).  The protocol is based on sending  from $\mathcal{A}$  product states of coherent states $\vert \alpha_{L1} \alpha_{S1}\alpha_{L2}\alpha_{S2}\rangle$ where $\vert \alpha_{Li}\rangle, i=1,2$ are local oscillators (to perform homodyne detection in $\mathcal{B}$) excited in a linear polarization mode, and  $\vert \alpha_{Si}\rangle$ are signal weak coherent states excited in the orthogonal polarization mode. All of them are  in the same quadrature ({\em basis}) and are propagated, for example, in two single mode optical fibres (SMF) or a two-core optical fibre.

The protocol  at the quantum lines  A-A$_{i}$, or B$_{i}$-B is as follows: for each single event, $\mathcal{A}$ (A or B) randomly chooses the same quadrature (or {\em{basis}}) $\mathcal{E}$ or $\mathcal{P}$ for all modes. If she chooses the first quadrature $\mathcal{E}$, she applies a random phase $\phi_k={0,\pi}$ to each mode. On the contrary, she applies a random phase
$\{ \pi/2, 3\pi/2 \}$ to select the second quadrature $\mathcal{P}$. Thus,  for $N=2$ the system $\mathcal{A}$   can  send the following eight states in the $\mathcal{E}_{1}\mathcal{E}_{2}$  and  $\mathcal{P}_{1}\mathcal{P}_{2}$ quadratures
\begin{equation}\label{estados+-}
\vert \alpha_{1} \, \alpha_{2} \rangle_{\mathcal{E}_{1}\mathcal{E}_{2}}= {\ket{\pm \,a \pm a }},
\end{equation}
\begin{equation}\label{estados+i-i}
%\vert \Psi_{\mathcal{P}_{1}\mathcal{P}_{2}}\rangle=
\vert \alpha_{1} \,\alpha_{2}\rangle_{\mathcal{P}_{1}\mathcal{P}_{2}}= {\ket{\pm ia\, \pm ia}},
\end{equation}
  with $a>0$. If $N=1$ then we would  have only the two states $\vert \pm a\rangle$,  in the $\mathcal{E}$-quadrature, and   $\vert \pm ia\rangle$ in the $\mathcal{P}$-quadrature. These four states were used in the seminal work by Namiki and Hirano \cite{Namiki2003} for  a single mode; however, for $N=2$ we have eight states, that is, $2^{2(=N)+1}$$=$$8$ states. If we have $N=3$ then  a third weak coherent state is considered, and thus we  will have the states $\vert \pm a  \pm a \pm a\rangle$,  in the $\mathcal{E}$-quadrature, and   the states $\vert \pm ia \pm ia \pm ia\rangle$ in the $\mathcal{P}$-quadrature (16 states), and so on for $N$ modes.    
  
  Afterwards,  $\mathcal{B}$ (A$_{i}$ or B$_{i}$)  randomly chooses to measure the first or the second quadrature by a  homodyne detection, that is, one of the  conjugated basis $\mathcal{E}/\mathcal{P}$ is selected  by applying the same phase $\phi_{B}=0$ or $\phi_{B}=-\frac{\pi}{2}$ to the local oscillator for every mode. If Bob  performs a homodyne measurement for all modes with  phase $\phi_{B}=0$ a  product of weak coherent states  in the optical field ${\mathcal{E}}$ basis for each mode will be measured, otherwise  if $\phi_{B}=-\frac{\pi}{2}$ a measurement in the optical field momentum ${\mathcal{P}}$ basis is made.
  
\begin{figure}[h]
\centering
     \begin{subfigure}[h]{0.4\textwidth}      
         \includegraphics[width=1.05\textwidth]{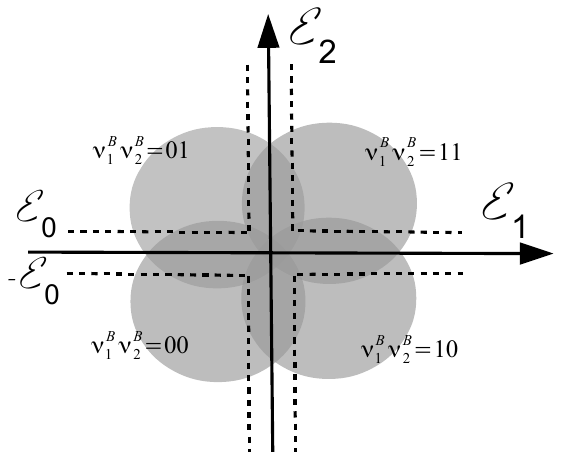}
         \caption{}
         \label{fig0a}      
     \end{subfigure}
     \hfill
 \begin{subfigure}[h]{0.4\textwidth}
            \,\hspace{-1.95cm} \includegraphics[width=1.25\textwidth]{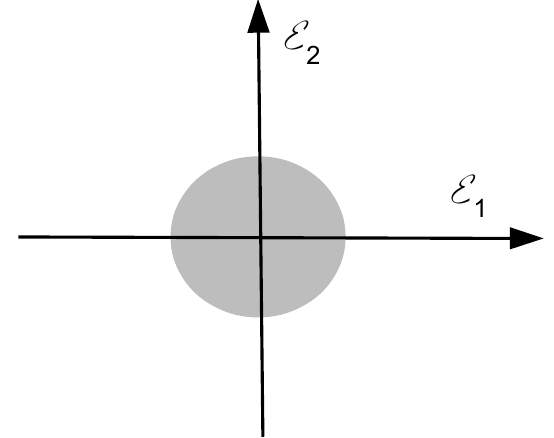}
         \caption{ }
         \label{fig0b}
     \end{subfigure}
   \caption{(a) The four weak coherent states for  $N$$=$$2$ modes represented in the two-dimensional first quadrature space $\mathcal{E}_{1}\mathcal{E}_{2}$. The states are centred on $(\mathcal{E}_{1}\mathcal{E}_{2})=(\pm a \pm a \rangle$.  The values $\nu_{1}^{B}\nu_{2}^{B}$ for $N$$=$$2$ are indicated  within the threshold regions with squared frontiers in the two-dimensional first quadrature space $\mathcal{E}_{1}\mathcal{E}_{2}$; the same thresholds $\pm\mathcal{E}_{o}$ has been chosen for each dimension. (b) The states in the basis $\mathcal{P}$, that is, the vacuum state $\vert 00\rangle$ is shown.}
\label{fig02}
\end{figure}

   At this point  is interesting to note that in the $\mathcal{E}$-representation, and by using appropriate optical-quantum units,  the following equivalence is obtained $\mathcal{E}=\mathcal{R}e(\alpha)$ \cite{schleich01, Loudon1983}. In this representation $\mathcal{E}_{1}\mathcal{E}_{2}$ the probability of these states is given by the general expression
  \begin{equation}
    P(\mathcal{E}_{1},\mathcal{E}_{2})=\vert\langle {\mathcal{E}_{1}\mathcal{E}_{2}} \vert \alpha_{1} \, \alpha_{2} \rangle\vert^{2}=\frac{2}{\pi}\,e^{-2(\mathcal{E}_{1}-\mathcal{R}e(\alpha_{1}))^{2}}\,e^{-2(\mathcal{E}_{2}-\mathcal{R}e(\alpha_{2}))^{2}},
\end{equation}
where $\mathcal{R}e(\alpha_{1})$ and $\mathcal{R}e(\alpha_{2})$ are mean optical fields, 
therefore we have  the following  probabilities for the  eight states when are measured in the basis $\mathcal{E}_{1}\mathcal{E}_{2}$ when Bob applies a phase phase $\phi_{B}=0$, 
\begin{subequations}
\begin{equation}\label{P}
    P(\mathcal{E}_{1},\mathcal{E}_{2})=\vert\langle {\mathcal{E}_{1}\mathcal{E}_{2}} \vert \alpha_{1} \, \alpha_{2} \rangle_{\mathcal{E}_{1}\mathcal{E}_{2}}\vert^{2}=\frac{2}{\pi}\,e^{-2(\mathcal{E}_{1}\pm a)^{2}}\,e^{-2(\mathcal{E}_{2}\pm a)^{2}}
\end{equation}
\begin{equation}\label{P'}
    P'(\mathcal{E}_{1},\mathcal{E}_{2})=\vert\langle {\mathcal{E}_{1}\mathcal{E}_{2}} \vert \alpha_{1} \, \alpha_{2} \rangle_{\mathcal{P}_{1}\mathcal{P}_{2}}\vert^{2}=     \frac{2}{\pi}\,e^{-2(\mathcal{E}_{1}^{2}+\mathcal{E}_{2}^{2})}.
\end{equation}
\end{subequations}
 where $\mathcal{R}e(\alpha_{1})=\mathcal{R}e(\alpha_{2})=\pm a$ are mean optical fields.  In Fig.\ref{fig02} it is shown for   $N$$=$$2$ a top view of these probabilities functions. Note that if Alice sends states given by Eq.(\ref{estados+-}) then Bob  can determine with a high probability which state was sent by Alice, that is, he detects with probability given by Eq.(\ref{P}); however, if Alice 
sends states given by Eq.(\ref{estados+i-i}) then he can not determine with a high probability which state was sent by Alice because Bob detects with probabilities given by Eq.(\ref{P'}). In this case if Bob would have changed the basis, that is, $\phi_{B}=-\frac{\pi}{2}$,  then states would have again the probabilities given by Eq.(\ref{P}).  In short, these are  the main quantum properties to get a secure key in continuous variable.  

Next, Alice $\mathcal{A}$ and Bob $\mathcal{B}$, by public announce of bases, retain the pulses with coincident basis obtaining a sifted key. Due to overlapping among these states, a threshold $\mathcal{E}_{o}$ for the optical field detected   has to be chosen by Bob to discriminate between states (postselection) and therefore a bit assignment criteria is constructed. Thus, for $N=2$ we have  

\begin{equation}
\label{criteria_bob}
\nu^{B}_{1}\nu^{B}_{2}=
    \begin{cases}
        1 1 & \text{if} \hspace{1.0mm} \mathcal{E}_{1}\geq \mathcal{E}_{o} \hspace{1.0mm} \text{and } \hspace{1.0mm}  \mathcal{E}_{2}\geq \mathcal{E}_{o} \\
        1 0 & \text{if} \hspace{1.0mm} \mathcal{E}_{1}\geq \mathcal{E}_{o} \hspace{1.0mm} \text{and } \hspace{1.0mm}  \mathcal{E}_{2}\leq -\mathcal{E}_{o} \\
         0 1 & \text{if} \hspace{1.0mm} \mathcal{E}_{1}\leq -\mathcal{E}_{o} \hspace{1.0mm} \text{and } \hspace{1.0mm}  \mathcal{E}_{2}\geq \mathcal{E}_{o} \\
          0 0 & \text{if} \hspace{1.0mm} \mathcal{E}_{1}\leq -\mathcal{E}_{o} \hspace{1.0mm} \text{and } \hspace{1.0mm}  \mathcal{E}_{2}\leq -\mathcal{E}_{o} \\
              \text{none} & \text{if} \hspace{1.0mm}\vert\mathcal{E}_{1}\vert< \mathcal{E}_{o}\hspace{1.0mm} {\rm or}   \hspace{1.0mm}    \vert\mathcal{E}_{2}\vert< \mathcal{E}_{o},
    \end{cases}
\end{equation}
where $\mathcal{E}_{1,2}$ are the results of Bob's measurements. If some optical field value does not fulfil the criteria then no bits are distilled.   These thresholds could have  been selected arbitrarily, but for a practical case rectilinear frontiers have been selected, as shown  in Fig.\ref{fig02}, and accordingly square thresholds regions are defined. 
Therefore, when Alice announces the basis, that is, the quadrature used,  then  
the  Bob's task will be to distinguish as best as possible between the four states of the first ($\mathcal{E}$) or the second ($\mathcal{P}$) quadrature. The criteria given by Eq.(\ref{criteria_bob}) is used to perform the mentioned task by using a balanced homodyne detection and thus a secret key can be sifted. \textcolor{black}{The final step will be a connection with classical signal between A$_{i}$ and B$_{i}$ by using several security physical layers as shown in sections 3 and 4.}

\subsection{Quantum Bit Error Rate and Secure Key Rates}
\begin{figure}[h]
\centering
     \begin{subfigure}[h]{0.48\textwidth}      
         \includegraphics[height=\linewidth]{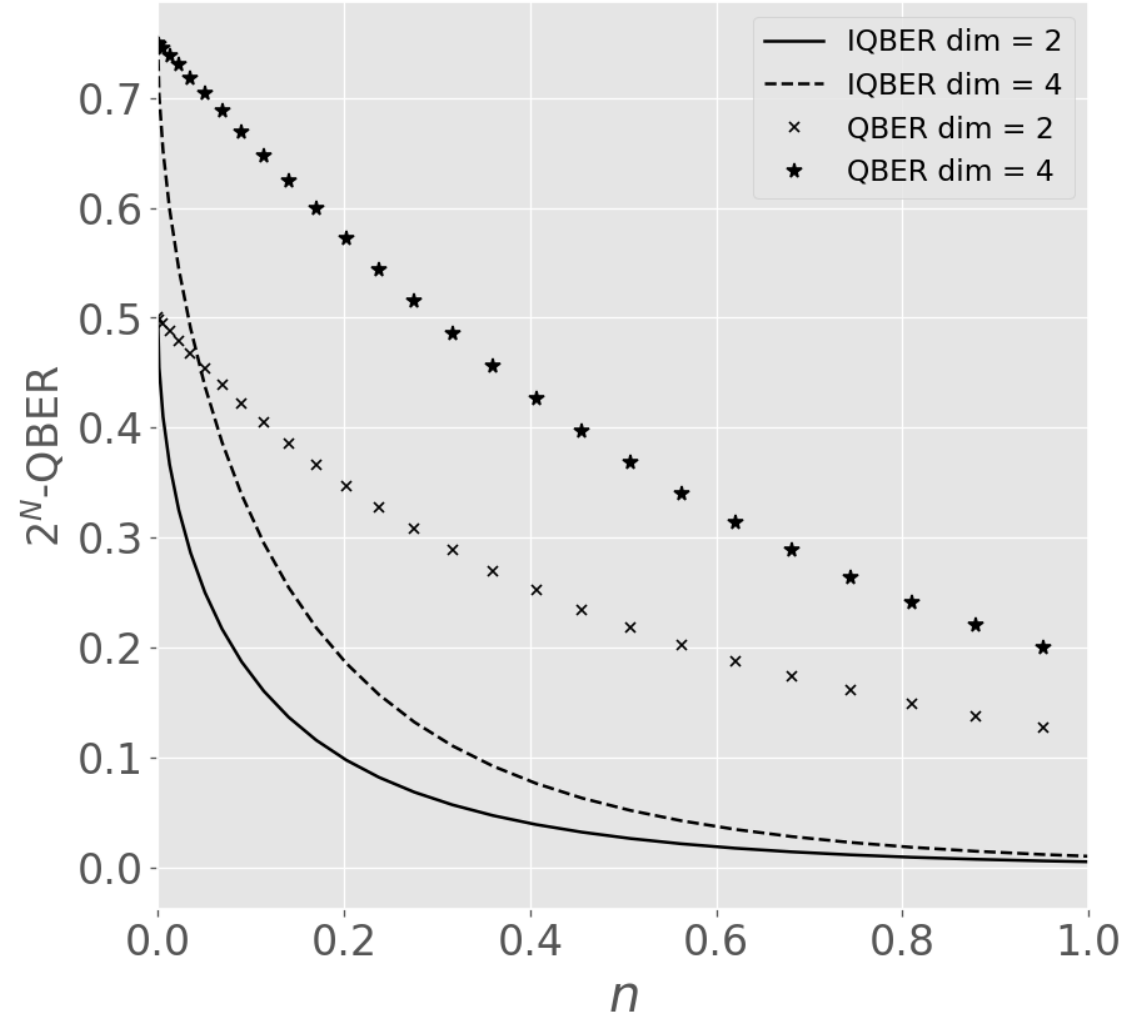}
         \caption{}
         \label{fig0a}      
     \end{subfigure}
     \hfill
 \begin{subfigure}[h]{0.51\textwidth}
         \includegraphics[height=0.985\linewidth]{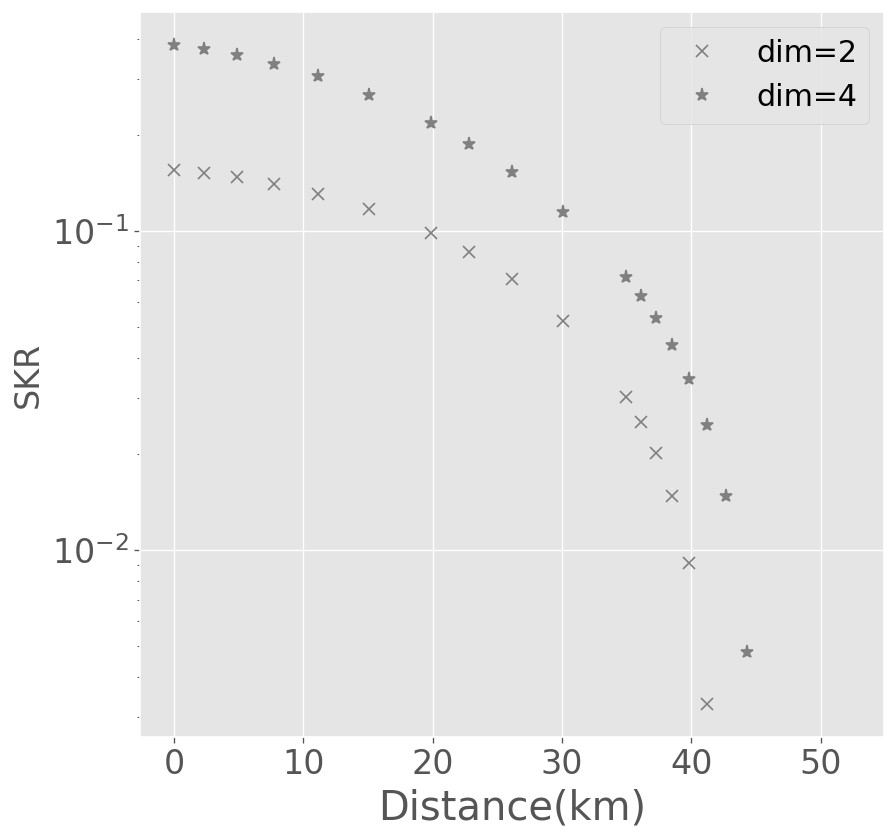}
         \caption{}
         \label{fig0b}
     \end{subfigure}
\caption{(a) QBER  for 1 and 2 modes (cross and star dots) as a function of mean photon number; IQBER  is shown for completeness (solid and dashed lines).  (b) Secret Key Rate (SKR) as a function of distance $L$ for 1 and 2 modes  (cross and star dots) for a mean optical field $a=1.2$ and threshold $\mathcal{E}_{o}=0.3$.}
\label{figQBER+SKR}
\end{figure}

\textcolor{black}{We present the Quantum Bit Error Rate (QBER) produced by both  homodyne detection and Eve's attacks along with the Secure Key Rates as a function of losses and therefore of distance $L$. All these results can be applied to the quantum lines A-A$_{i}$ and B$_{i}$-B}. We start with the homodyne detection that produces an Intrinsic Quantum Bit Error Rate (IQBER) which can be  eliminated by  error corrections. The intrinsic error consist of measuring, for example, $(\mathcal{E}_{1}>\mathcal{E}_{o}, \mathcal{E}_{2}<-\mathcal{E}_{o})$ or   $(\mathcal{E}_{1}<-\mathcal{E}_{o}, \mathcal{E}_{2}>\mathcal{E}_{o})$, or  $(\mathcal{E}_{1}<-\mathcal{E}_{o}, \mathcal{E}_{2}>\mathcal{E}_{o})$, and nevertheless Alice sent the state $\vert a a\rangle$ and therefore Bob should have measured  $\mathcal{E}_{1}>\mathcal{E}_{o}, \mathcal{E}_{2}>\mathcal{E}_{o}$. On the other hand, we will have QBER due to Eve's attacks, for example, an intercept-resend attack  by performing simultaneous measures in the two quadratures, therefore, Eve only can use the half of mean photons of each weak coherent state to identify the bits. \textcolor{black}{The corresponding QBER  and Secure Key Rate (SKR) has been  formulated and extensively studied by Namiki and Hirano for $N=1$ mode \cite{Namiki2003}, and recently,  the QBER and the SKR for HD-DM-CV-QKD ($N>1$ modes)   have been presented \cite{optik2025} where the SKR is clearly increased.}  In Fig.\ref{figQBER+SKR}, and for an illustrative purpose, we present results for $N=1,2$ of the  IQBER and QBER  as a function of the mean number of photons and  for a fixed threshold $\mathcal{E}_{o}=0.3$.  Likewise,  for $N=1,2$ modes and  a Eve's partial attack of 50\% (fraction of attacks $\eta=0.5$), the  \textcolor{black}{SKR is plotted in   Fig.\ref{figQBER+SKR}   as a function of the distance $L$ for an optical fibre attenuation $\alpha_{att}\approx 0.2$ dB/km, a mean optical field  $a=1.2$ and a threshold $\mathcal{E}_{o}=0.3$. %\textcolor{black}{Losses are in turn related to  the propagation distance  $L$ depending on the optical fibre attenuation $\alpha_{att}$, that is, $L=(10/\alpha_{att})\log(1-{\rm Loss})^{-1}$ \cite{PRA21}. 
Note that  QBER increases with $N$ and, therefore,  so does security, and besides the SKR takes reasonable values under Eve's attack.  The most important conclusion is that the SKR increases with dimension, and therefore we can use a greater QKD distance $L$.  Thus, distances around $L\approx 45$km are obtained, therefore, the lines A-A$_{i}$ and B$_{i}$-B should  be located on narrow continental shelves.}  On the other  hand, it is also interesting to indicate that the rate $R_{2}$ for $N=2$ is greater than the corresponding to two independent single channels with rate $R_{1}$, that is, $R_{2}>2R_{1}$. In these calculations we have assumed that errors due to optical perturbations are negligible because an active or passive compensation technique has been applied to the perturbed quantum states, as shown in the next subsection.

\subsection{Efficient QKD  and reduction of technological complexity by autcompensating} 

\begin{figure}[h]
\centering
\includegraphics[width=0.81\textwidth]{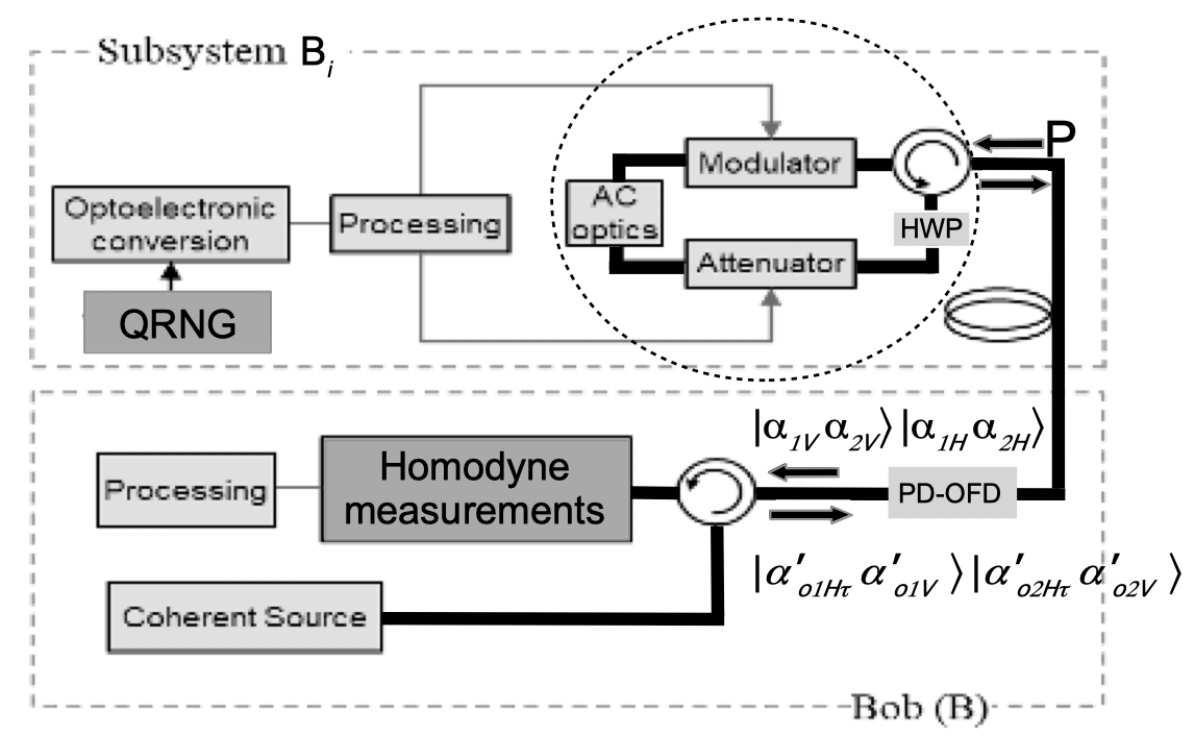}
\caption{Sketch of the autocompensation system. It is shown a product  coherent state with two polarizations one for the local oscillator and the other one for the signal. An arbitrary perturbation P is indicated. Note the location of the HWP (Half Wave Plate) in the closed cycle (enclosed in a dotted circle) and the location of the PD-OFD (Polarization Dependent Optical fibre Delay) at the start of the quantum line.}
\label{Auto1}
\end{figure}

It is well-known that quantum states are perturbed under propagation in optical fibres due to intrinsic imperfections, mechanical and thermal perturbations and so on. Accordingly,  errors appear and QKD efficiency is reduced in the quantum lines, and  therefore it is important to restore the quantum states. For that, active compensators of the perturbations can be used although present a certain technological difficulties. Alternatively, passive compensation techniques (autocompensation) can also be used  \cite{PLUG97, Bethune02, Kawamoto2005, Bal19,  PRA21}. 
We must note that these passive techniques have been proposed to achieve a plug and play system and thus  to increase the key-rate and transmission distance, that is, the efficiency of the QKD protocols. They use a round-trip propagation that compensates, by inserting a proper modal unitary transformation, the perturbations undergone  by the quantum states.   In our case, A and B will play the role of $\mathcal{B}$  and A$_{i}$ and B$_{i}$ the role of  $\mathcal{A}$, and     the autocompensating technique will be applied to polarization modes and  implemented by locating the sources and homodyne detectors at terrestrial points A and B,  which has a certain advantage as to reduce the technological complexity.   On the other hand, quantum random number generators (QRNG) and phase modulators will be located at the submarine subsystems A$_{i}$ and B$_{i}$.   As commented, an active compensating device could be  used, but a  passive compensation (autocompensation) of QKD supposes a very practical solution.  Thus, optical elements like Faraday Mirrors can be used to implement autocompensation in polarization modes, but in  our case we will use a Half Wave Plate (HWP) which, as we will show, is a most versatile  solution to implement an autocompensating HD-DM-CV-QKD protocol.   
First of all, we present the fundamentals   to understand how a HWP implements autocompensation in the polarization modes by using a closed cycle located, for example,  in subsystem B$_{i}$ belonging to the arbitrary quantum line B$_{i}$-B. In  Fig.\ref{Auto1}  a sketch of the autocompesation system is shown (optical fibre lines  are  indicated with thick lines and electrical lines with  thin lines).  At the top right, an optical fibre  closed cycle  is shown (enclosed in a dotted circle) where the input to the cycle and output from the cycle is implemented by an Optical Circulator (OC). The cycle contains  Modulators and  Attenuators controlled by AC optics, and a HWP.
 
 Let us consider an anisotropic perturbation P before the closed cycle, that is, a perturbation of the polarization of the quantum states. An anisotropic perturbation is represented by an unitary matrix $M$ for progressive modes and $M'$ for regressive modes, that is, 
 \begin{equation}
 M=\begin{pmatrix} a & ib \\ ib & a^{*}  \end{pmatrix}, \quad M'=\begin{pmatrix} a & -ib \\ -ib & a^{*}    \end{pmatrix},
\end{equation}
where parameter $b$ is a real number, then, by using a HWP rotated $\pi/4$ an antidiagonal  matrix $D'$ is implemented and then we obtain
\begin{equation}
  M'D'M= \begin{pmatrix} a & -ib \\ -ib & a^{*}    \end{pmatrix} \begin{pmatrix} 0 & 1 \\ 1 & 0  \end{pmatrix}\begin{pmatrix} a & ib \\ ib & a^{*}    \end{pmatrix}= \begin{pmatrix} 0 & 1 \\ 1 & 0  \end{pmatrix},
\end{equation}
that is, an autocompensation is obtained although polarizations have been permuted which has not any relevant effect. Obviously, if there are more perturbations the same autocompensation is obtained. In short, if we start with a coherent state $\vert \alpha_{H}\alpha_{V}\rangle$, the state after a round trip,  where it has  undergone perturbations,  becomes  the   state $\vert \alpha_{V}\alpha_{H}\rangle$. This will be the result that we will use to get autocompensation (technical details  can be found in \cite{Bal19,PRA21}). 
 
 Now we explicit a little more the physical process in any subsystem B-B$_{i}$ (or   A-A$_{i}$) to achieve an autocompensating QKD protocol. The laser (coherent) sources and detectors (homodyne detection connected to a processing unity) will be located at a terrestrial point B as shown in Fig.\ref{Auto1}, which offers a certain advantage, as commented. 
 We have to remember that the protocol described above is  implemented by a  scheme assisted by polarization, that is, quantum states (signal) and  LO will be excited in different polarization modes, and moreover several types of fibres can be used such as few mode fibres (FMF), multicore fibres (MCF) and even a bundle of single-mode fibres (SMF).  In all these cases  it can be  assumed  that there is no  spatial coupling, particularly if we use SMF, however, polarization  modal coupling limits their applicability for long distance links. Such coupling is due to slow external random perturbations (mechanical, thermal...) or to waveguide birefringence (imperfections) of the optical fibres, and thus an autocompensation of these perturbations is required.

Autocompensation is obtained as follows:  Bob emits a spatial  two-mode coherent pulse  with the same mean number of photons in each mode (we must indicate that now the role of Alice $\mathcal{A}$ is played by  B$_{i}$ and the role of Bob $\mathcal{B}$  is played by the own B). The above two-mode coherent pulse can be obtained by using a coherent source (a laser), as shown in Fig.\ref{Auto1}, which can be divided by a beam splitter and coupled to two SMF optical fibres. 
The output state from Bob can be written as follows
\begin{equation}
\vert \Psi_{o}\rangle= \vert \alpha'_{o1H\tau}   \alpha'_{o1V} \rangle \, \vert  \alpha'_{o2H\tau} \alpha'_{o2V}\rangle,
\end{equation}
where $\tau$ is a delay produced  by a Polarization Dependent Optical fibre Delay (PD-OFD \cite{Bal19} and   it has been assumed that $\alpha'_{oiH}=A'e^{i\phi_{oi}}, \alpha'_{oiV}=A'e^{i\phi_{oi}}$ ($i=1,2$), that is, the coherent states in each polarization space have  the same phase $\phi_{oi}$.  

Next, under propagation each coherent state is perturbed, for example, $\vert \alpha'_{o1H}\rangle\rightarrow  \vert \tilde{\alpha}_{o1H} \tilde{\alpha}_{o1V}\rangle$. When the signal arrives to Alice, that is, to the mentioned closed cycle at B$_{i}$,  each perturbed polarization coherent state of the product state passes through a  HWP with angle $\pi/4$ (HWP$_{\pi/4}$)  and therefore  compensation of polarization perturbations is achieved when the pulses return to Bob, as analysed above, although polarization modes will arrive permuted (H$\rightarrow $V, V$\rightarrow$H) (that is, a logic X gate is applied). Moreover, in such a closed cycle a modulation (determined by a QRNG and an optoelectronic conversion and the corresponding processing) and an attenuation of the signal product coherent state is realized. In short, an autocompensation is successfully performed  and the  state reaching the detection system at Bob is given by
\begin{equation}
\vert \Psi\rangle=\vert \alpha_{1V}  \alpha_{1H}  \rangle\, \vert \alpha_{2V}   \alpha_{2H} \rangle,
\end{equation}
which, depending on the (random) modulation and the attenuation applied in the closed cycle, are given by the following expressions
\begin{equation}
\vert \Psi\rangle=\vert \pm a e^{i\phi_{1}}\ A e^{i\phi_{1}}    \rangle\, \vert \pm a e^{i\phi_{2}}\ A e^{i\phi_{2}}\rangle
 \quad \mathcal{E} \text{ basis}
\end{equation}
or 
\begin{equation}
\vert \Psi\rangle=\vert \pm ia e^{i\phi_{1}}\ A e^{i\phi_{1}}   \rangle\, \vert \pm ia e^{i\phi_{2}}  \ A e^{i\phi_{2}}\rangle \quad \mathcal{P} \text{ basis},
\label{pi2}
\end{equation}
where $a$\,$\ll $\,$A$\,$<$\,$ A'$, that is, $\vert \alpha_{1V}  \alpha_{2V}  \rangle$ is a weak coherent state, and $\phi_{1}$ and $\phi_{2}$ are global phases acquired by the optical modes without any relevance in the homodyne detection.  The final measuring step (homodyne measurements) begins with a PBS at $45^{o}$ which splits again the light into signal and LO \cite{Namiki2003, optik2025}, both polarized at the contrary as Bob's output, due to the permutation of the HWP. For each spatial mode an Electro-Optical Phase-Modulator is used, which applies a random phase $\phi_{B}$ to the LO (horizontal polarization) and  the same for the LO of  each mode. Finally, a homodyne detection is made at each spatial mode by using  a pair of photodetectors capable of measuring the homodyne current. In short, the HD-DM-CV-QHD protocol is  implemented and a high rate of secure key can be distilled.

\subsection{Hybrid HD-DM-CV-QKD protocol for the entire  subsea  line A-B}\label{HProt}
In the subsection \ref{2-1} the roles of $\mathcal{A}$-$\mathcal{B}$ were played by A$_{i}$-A,  and B$_{i}$-B, that is, in each quantum line a key is obtained.  Thus in quantum line A- A$_{i}$ a key k$_{A_{i}}$ has been obtained, then A$_{i}$ can send this key through the non quantum  channel between A$_{i}$ and B$_{i}$, that is, by using classical signals, which will be protected with several security physical layers. At  B$_{i}$ a key k$_{B_{i}}$ has been obtained in the  quantum line B$_{i}$-B that can be used as $k_{A_{i}} \oplus k_{B_{i}}$ to send it to B. In short, A and B will share a key k$_{A_{i}}$. Note that his scheme requires two QKD systems. 

On the other hand, it would be possible  to reduce the technological complexity by using only one QKD system A-B,  where A$_{i}$ and B$_{i}$  are intermediate points where modulation is performed, and the lines A$_{i}$-B$_{i}$  are non quantum  ones once more.  We briefly describe this hybrid protocol  \cite{Suboptic1}, \textcolor{black}{in which we  again consider the autocompensating system explained in the above section}. Let us consider that at A$_{i}$ there is a QRNG  device which determines in a random way what states must be sent to A and to B from A$_{i}$. Thus, the QRNG determines  the phase  modulation ($0,\pi,\pi/2;3\pi/2$), according to the DM-CV-QKD protocol explained in subsection  \ref{2-1}, that  must be applied at each mode of a strong monomode  (or multimode) coherent state  coming from A, and redirected to A as a quantum state obtained by attenuation to a weak coherent state level. On the other hand,  the information will reach B$_{i}$ by a non quantum channel but in deep waters \textcolor{black}{(natural protection), that is, the above phase modulations are coded 
in classical signals propagating from A$_{i}$  to B$_{i}$. This classical signal will indicate at B$_{i}$   which phase modulations  have to be made to generate the quantum states that have to be send to B, that is, in the autocompensating configuration which phase modulation   must be applied to a strong monomode  (or multimode) coherent state  coming from B, and redirected to B as a quantum state obtained by attenuation to a weak coherent state level. } In short, after modulation the states are attenuated and sent from A$_{i}$ to A,  and from   B$_{i}$ to B.   At  A and B homodyne detection is used to measure the quantum  states  (weak coherent states). For that,  A and B have to choose  which basis to use, that is,  first or second quadrature, to measure the quantum state. Next, by  public channel, A and B announce the bases used,  and moreover  A$_{i}$ also communicate the basis of each of   the states sent. Note that in this case there are bases coincidences in   25\%, unlike the case non hybrid of section  \ref{2-1} where the bases coincidences was 50\%. With this information a key can be shared between A and B. Note that the Eve's attack has also  to be made  in the non quantum lines A$_{i}$-B$_{i}$, but now the QRNG systems  are reduced by half, which, once more allows a reduction of the technological complexity. \textcolor{black}{Note that the QBER will be the same that the studied in subsection 2.3 for attacks in the quantum sections of the line (A-A$_{i}$ and B$_{i}$-B),  and SKR will be reduced by half compared to the protocol on the continental shelf.
}

\section{Hybrid approach with M combined submarine channels}

Submarine optical communications are usually established between points A and B separated a long distance. The standard hybrid solution would be the use of several submarine trusted nodes which supposes a high technological difficulty since the current submarine lines have not these nodes for quantum communications. Obviously, the region between, for instance, A$_{1}$-B$_{1}$ (see Fig.\ref{Figinicial}), could be considered a node but not fully trusted, because although an optical attack, such as optical fibre tapping, would have an extremely high technological difficulty, it would be possible. Therefore, the solution would be to add one or several physical layers of security. 
\begin{figure}[htbp]
\centering\includegraphics[width=10cm]{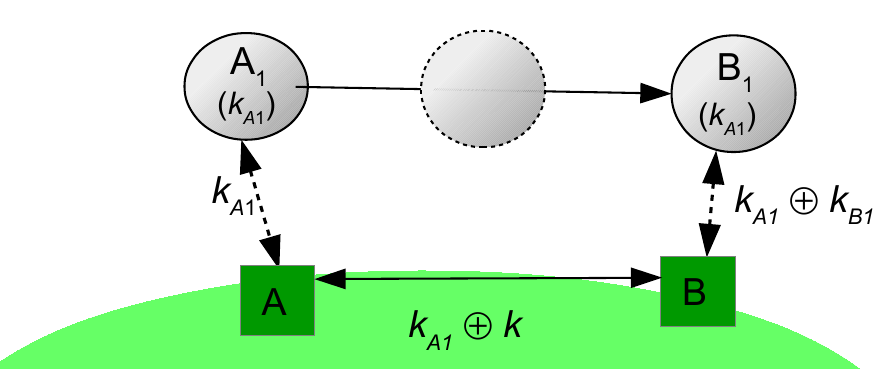}
\caption{Sketch of extra physical layer of security  between A and B connected with one satellite as trusted node travelling from a point above A to a point above B.}
\label{S1}
\end{figure}

We must stress that a similar situation is found in Earth-Satellite QKD  where the region A$_{1}$-B$_{1}$ would correspond to the satellite (see Fig.\ref{S1}) which contains a classical signal that can be copied, that is, there is no quantum security within the satellite \cite{Bedington2017, Liao2018}, therefore satellite   must be considered as a flying-trusted node.  
As suggested by Huttner {\em et.al.} \cite{Huttner2022} an extra physical layer security can be included by introducing a second channel, in this case a terrestrial one, apart from  the first channel (satellite channel), and then two secret  keys are combined and transmitted over two channels.

We will adapt this suggestion to submarine optical communications. We briefly remember this hybrid approach for Earth-Satellite QKD. In Fig.\ref{S1} a satellite shares a key $k_{A1}$ at the quantum line A-A$_{1}$ and another key $k_{B1}$ at the quantum line B$_{1}$-B. The satellite transports the key  $k_{A1}$  from a point over A to a point over B and then it  sends the key  under the XOR operation $k_{A1}\oplus k_{B1}$. Obviously, the satellite is a classical channel, therefore has not a fundamental protection, it only has a certain physical protection corresponding to the difficulty for accessing to it, and then the best option is to include an extra physical layer of security. Such as  suggested in \cite{Huttner2022} a classical terrestrial channel can be used where a new key $k$ is transmitted as $k_{A1}\oplus k$ from A to B, and thus Eve has to break security in two channels; we must note that  the attack can be made in any point of the terrestrial optical line.

\subsection{Double radial multichannel topology}

\begin{figure}[htbp]
\centering\includegraphics[width=9.5cm]{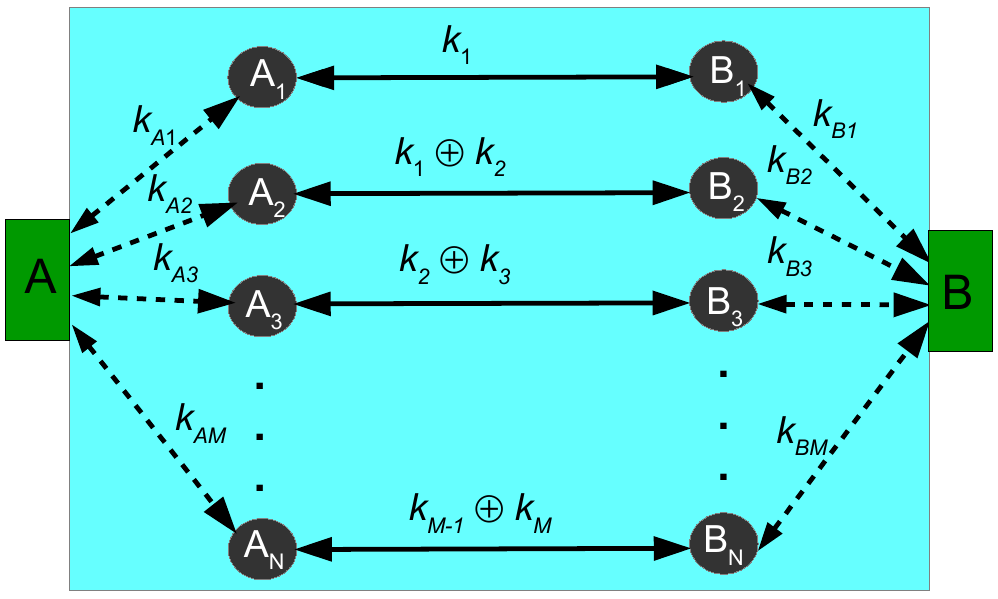}
\caption{Sketch of the double radial topology of $M$ submarine (hybrid) channels between two nodes A and B.}
\label{FigH1}
\end{figure}

The above hybrid approach for earth-satellite-earth will be adapted  to the \textcolor{black}{hybrid} submarine case, that is, \textcolor{black}{the extra optical line(s) (that  provide additional security physical layers)} should be located as far  into the sea as possible to increase the difficulty of an optical attack. Let us consider the  additional points A$_{i}$ and B$_{i}$, $i=1,...,M$, therefore there are $M$ channels (multichannel scheme) between A and B (double radial multichannel topology), that is, A is connected  to $M$ subsystems A$_{i}$ and B is connected  to $M$ subsystems B$_{i}$ with quantum lines, such as shown in Fig.\ref{FigH1}, that is, in each quantum line a key is shared, that is, $k_{Ai}$ and $k_{Bi}$.  The lines A$_{i}$-B$_{i}$ are classical optical lines far from the coast, as shown in  Fig.\ref{Figinicial}.
With this topology we can generalize the above approach  \textcolor{black}{used for earth-satellite-earth}  to  $M$ channels,  and moreover we will be able to  introduce different combinations of keys.  

 \textcolor{black}{We will use the protocol described in subsection 2.1, that is, the quantum lines A-A$_{i}$ and B$_{i}$-B, located on the continental shelf, generate quantum keys. We must indicate that the hybrid protocol described in subsection 2.4 could be also used. } We start with the first key $k_{A1}\equiv k_{1}$ generated in the quantum line A-A$_{1}$ 
\textcolor{black}{(this is analogous to the path A-A$_{1}$ in the satellite case, as shown in Fig.\ref{S1})}; 
 \textcolor{black}{next, this key $ k_{1}$  is sent along the non quantum channel 1 from A$_{1}$ to B$_{1}$. } \textcolor{black}{Note that this is analogous to the path  A$_{1}$-B$_{1}$ in the satellite case, as shown in Fig.\ref{S1}, that is, the submarine non quantum channel sends the key by using classical signals and therefore optical amplifiers can be used, which is analogous to  the classical transport of the quantum key inside the satellite from A$_{1}$ to B$_{1}$. 
Finally, at   B$_{1}$ a quantum key $k_{B1}$ is used to forward to B the XOR of  key  $k_{1}\equiv k_{A1}$ (this is analogous to the quantum line   B$_{1}$-B in the satellite case, as shown in Fig.\ref{S1}).  
 Next, a first additional physical security layer is introduced. The key $k_{A2}$ is used in A to forward to A$_{2}$ the  XOR of this key with a composed key $k_{2} \oplus k_{1}$, that is,  $k_{A2}\oplus (k_{2} \oplus k_{1})$, where  the key $k_{2}$ is generated at A (note that $k_{A2}=k_{2}$ could also be chosen). Next,  the key $(k_{2} \oplus k_{1})$ is sent along channel 2 to  B$_{2}$ by classical signals. At point B$_{2}$ the composed  key $(k_{2} \oplus k_{1})$ is transmitted by the quantum channel to B by making a XOR operation with  the key $k_{B2}$. At this stage the key shared between A and B is $k_{2}$, therefore  with the additional channel  A$_{2}$-B$_{2}$ we have introduced an important layer of security, analogous to  the additional classical terrestrial line AB  of the  satellite-earth  case (see Fig,\ref{S1}). However we  emphasize that in our case the classical lines are as far  into the sea as the quantum lines allow, that is, the additional line has two quantum parts AA$_{2}$ and B$_{2}$B,  that is, the analogy with  the satellite-earth case would be to use a second satellite.}

The security can be  increased by  introducing an arbitrary number $M$ of these hybrid channels according to both the existing fibre optic network and the technological possibilities. Thus the key $k_{A3}$ is used  in A to forward to A$_{3}$ the  XOR of this key with a new composed key $k_{3}+k_{2}$ (the key $k_{3}$ is generated at A), which  is sent along channel 3 to  B$_{3}$.   This composed  key is transmitted to B by making a XOR operation with  the key $k_{B3}$. At this stage the key shared between A and B is $k_{3}$, therefore  with this additional channel  A$_{3}$-B$_{3}$ an additional layer of security is introduced in deep waters. In short, by this concatenated local processing, that is, a XOR in each channel, the key shared between A and B will be $k_{M}$, therefore Eve must  attack $M$ classic lines located  under the sea and separated from each other by a distance of several kilometres.

Alternatively, a simpler approach can be implemented if  a key  is constructed by combining the quantum keys $k_{Ai}$, that is, by choosing a common logic operation in A and B the new key is obtained.  We can use  a  XOR logic operation, a XNOR logic operation or combinations of them (or other operations). \textcolor{black}{For example the final key can be given by %  therefore we could obtain 
\begin{equation}\label{sumabin}
k=k_{A1}\oplus k_{A2}\oplus ... \oplus k_{AM-1}\oplus k_{AM}, 
\end{equation}
}
\textcolor{black}{The quantum keys $k_{Ai}$ (received in A$_{i}$) are sent along the non quantum lines (that is, in a classical way)} to subsystems B$_{i}$ where the operations XOR  $k_{Ai}\oplus k_{Bi}$ are made and sent to B. Since A and B will make the same logic operations with the keys  received a new key will be shared between them.   The main advantage of this  approach is that \textcolor{black}{both key generation at A and performing logical operations at A$_{i}$ are not required.}

% it does not require to generate keys at A and  do  logic  operations  in nodes A$_{i}$.  

\subsection{Non-radial multichannel topology} 

\begin{figure}[h]
\centering\includegraphics[width=9.5cm]{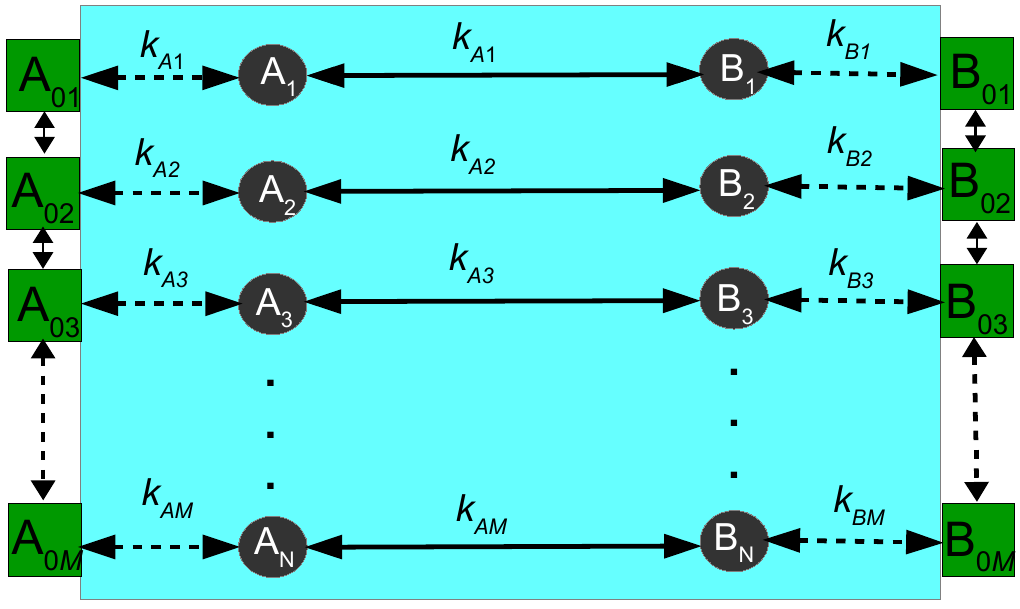}
\caption{Sketch of the non radial topology of $M$ submarine (hybrid) channels between  nodes A$_{0i}$ and B$_{0i}$.}
\label{H2}
\end{figure}

A most flexible topology can be used as shown in Fig.\ref{H2}, that is, we consider $M$ trusted nodes  A$_{oi}$ outside  the sea and connected between them by quantum  lines, where any of them can be Alice, and, on the other side of the sea, we will also have  $M$ trusted nodes B$_{oi}$ outside  the sea and also connected by quantum lines, where any of them can be Bob. Therefore this non radial topology would require to place  trusted nodes in   A$_{oi}$ and   B$_{oi}$ or to use the existent standard terrestrial nodes;  besides the quantum lines  would go straightforwardly  into   the sea to the nodes A$_{i}$ and B$_{i}$ without following oblique paths as in the above radial topology,  and accordingly the distance to the  subsystems located  further away from A (or B) would be reduced. Likewise, as commented,  with this topology the systems A and B could be any terrestrial trusted node  A$_{0i}$ and B$_{0i}$ providing a greater flexibility to the system, although  all the keys $k_{Ai}$  have to be sent to the trusted node playing the role A, and all keys $k_{Bi}$ have to be sent to the trusted node playing the role  B. Obviously this is a  small  disadvantage compared to the double radial topology where all signals is going out from A to B.

On the other hand, we take advantage of this topology to indicate how to apply the hybrid protocol described in section \ref{HProt}. Such a protocol would be implemented in each channel A$_{0i}$-B$_{0i}$, therefore we have the keys $k_{Ai}=k_{Bi}\equiv k_{i}$ which are protected in a quantum way in the paths A$_{0i}$-A$_{i}$ and B$_{i}$-B$_{0i}$, and by the multichannel hybrid scheme approach in deep waters. As commented, with the hybrid protocol we reduce the number of QRNG devices, and moreover the keys $k_{Bi}$ are not necessary, therefore the  multichannel hybrid approach is simplified. Obviously the double radial topology could also be used with this hybrid protocol.

In short, $M$ extra physical layers of security have been introduced.  We must stress that $M$ submarine optical fibre channels would have to be hacked and besides to establish an efficient communication system between hacked points On the other hand, we could  even introduce  Additional Physical Layers of Security (APLS) in a part of the non quantum lines. In the next section some  APLS  will be presented, many of them  are related to the type optical fibres used in the channels. 

\section{Hybrid approach with additional  physical layers of security}

We present in this section several APLS in an increasing order of technological complexity. These layers can be implemented along an optical line between  A$_{i}$ (or B$_{i}$) and some point  A'$_{i}$ (or B'$_{i}$)  beyond continental shelf, if possible,  that is, along the buried segment of the submarine optical communications lines as shown in Fig.\ref{Figinicial}.  We have to indicate that the arbitrary point A'$_{i}$  could reach the point B$_{i}$ if it were technologically feasible. Therefore,  optical lines   would be protected by additional layers of security in submarine depths less than 1000m,  which would require  Eve to perform optical tapping  in  much deeper waters.  We will present these APLS into two categories:  passive APLS and active APLS, if moreover an additional optoelectronic processing is required.

\subsection{Hybrid protocol  with passive APLS}
APLS can be introduced by  taking into account  both the  properties of the optical fibres used in the lines A$_{i}$-A'$_{i}$ (or B'$_{i}$-B$_{i}$) and  the protocol QKD used in the quantum lines. In section 2 we have presented a protocol with $N$ modes in each quantum line. This multimode solution is highly compatible with both single mode fibres and multicore optical  fibres. Therefore, we present some passive physical layers of security implemented with such optical fibres.

\subsubsection{APLS with $N$ single mode optical fibres}
Let us consider the general case of $N$ modes at the generic submarine quantum lines A-$A_{i}$, then $N$ bits can be distributed in $N$ modes in each quantum line.  Next, these bits are coupled to $N$ single mode fibres at each node $A_{i}$. Therefore, Eve will have to tap a number $N$ of SMFs (optical lines) at A$_{i}$-A'$_{i}$ of each submarine channel and therefore a total number $MN$ of SMFs. This represents a straightforward  physical layer of security by using optical fibres which could already be installed in the classical channels.

\subsubsection{APLS with multicore optical fibres}
On the other hand, if we have multicore optical fibres then new physical layers of security can be incorporated. Let us consider, for the convenience of exposition, $N=2$ (two bits) at the submarine quantum line A-$A_{1}$, then each bit  can be  transmitted   in a different  spatial mode. Next, these bits are coupled to a two-core optical fibre located at the classical channel segment A$_{1}$-A'$_{1}$. A tapping on  this optical fibre can be made, however Eve has to distinguish which core each bit originates from.  Obviously for several cores the difficulty is increased. Therefore, tapping has to be completed with some optical device that can determine from which core the light goes out. Obviously, if these two-core optical fibres are already installed in all submarine optical lines (or some of them) this passive APLS  could already  be introduced.

\subsubsection{APLS with spatial and/or polarization  modal coupling}
We can use the spatial or polarization modal coupling to protect the bits information. Thus, we can  transmit each bit  in a different  polarization mode, then modal coupling appears and therefore Eve has to implement some optical device to restore information (MIMO, ...) \cite{Bai2012}. The disadvantage in this case is that the optical devices  to restore information have to be also installed at one or several  submarine channels  A$_{i}$-B$_{i}$.

\subsection{Hybrid protocol  with  active APLS}
The above physical layers are based on the technological difficulty to make tapping and extract information, however we can also  use the own optical fibres to detect attacks by optical  tapping, that is, by processing the  optical signals.  We present two active physical layers with a certain detail, the first one is based in a simple detection of energy in a  dark mode, and the second  one is based in the use of a secret key obtained by classical perturbations (therefore it has not a fundamental protection). Besides, these active APLS can be mixed with the above passive ones.

\subsubsection{Active APLS based on modal power transference}
Let us consider, for example, optical fibres with three cores, two of them are single-mode cores and one of them is a few-mode core. Tapping can be considered as an optical perturbation and can therefore induce modal coupling. If the fundamental mode is excited in the few-mode core, a fraction of optical power can be transferred to the first excited mode by tapping, allowing the presence of Eve to be detected at the fibre end by measuring the energy in the first excited mode.  A perturbative theory of modal coupling can be used to calculate the energy transferred from the fundamental mode \cite{alex2025} to other modes.  Next, we present a possible modelling of the perturbation by optical tapping by fibre bending and the modal coupling efficiency from fundamental mode to the first excited mode in a Few Mode optical Fibre (FMF) (in our case a few-mode core).
For illustrative purposes we assume that the parabolic approximation for the first modes of a graded-index optical fibre can be made, that is, the refractive index profile is given by 
 \begin{equation}
n^{2}(x,y)\approx  n_{o}^{2}[1-g_{o}^{2}(x^{2}+y^{2})],
\end{equation}
where $n_{o}$ is the axial index and $g_{o}$ the graded-index parameter, therefore, separable Hermite-Gaussian modes will be used for numerical calculations. First of all we write the expression of the coupling coefficients between modes of order $m$ and $n$ under a perturbation $\Delta n^{2}(x,y,z)$,
\begin{equation}
K_{nm}=\omega\epsilon_{o}\int_{-\infty}^{+\infty}\int_{-\infty}^{+\infty}\mathrm{d}x\mathrm{d}y \Delta n^{2}(x,y,z)e_{n}(x,y)e^{\star}_{m}(x,y),
\end{equation}
where $\omega$ is the angular frequency, $\epsilon_{o}$ s the vacuum permittivity, and $e_{n} (x,y)$, $e_{m} (x,y)$ are the normalized optical modes \cite{Lee} of order $n$ and $m$, respectively, with propagation constants $\beta_{n,m}=k_{o}N_{n,m}$, where $N_{m,n}$ are the effective indices.  Next, by taking into account the above coupling constants and the propagation constants, the well-known asynchronous modal coupling equations system $-i\mathrm{d}a_{n}/\mathrm{d}z= a_{n}+K_{nm}a_{m}$, with repeated indices indicating sum, has to be solved for the mode amplitudes $a_{n}=A_{n}e^{i\beta_{n}z}$. Then under the perturbative approximation we have the following formal solution for the modal amplitudes $A_{n} (z)$ at the end of the fibre ($z$-direction) 
\begin{equation}\label{pert-coupling}
A_{n}(z)\approx\sum_{n}\int_{-\infty}^{+\infty} \mathrm{d}z\,A_{m}(0)K_{nm}(z)e^{i(\beta_{m}-\beta_{n})z}.
\end{equation}
We have also  assumed that the initial amplitudes $A_{n}(0)$ can be considered as constants in the perturbative approximation. 

\begin{figure}[h]
\centering
     \begin{subfigure}[h]{0.4\textwidth}      
         \includegraphics[width=1.05\textwidth]{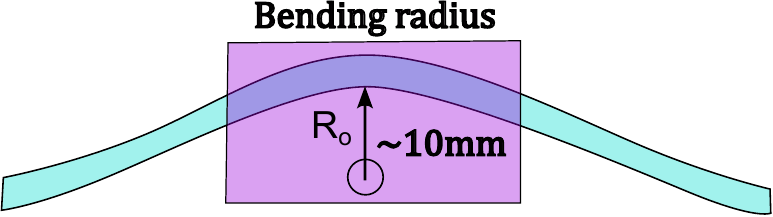}
         \caption{}
         \label{figtap-a}      
     \end{subfigure}
     \hfill
 \begin{subfigure}[h]{0.4\textwidth}
            \,\hspace{-1.95cm} \includegraphics[width=1.25\textwidth]{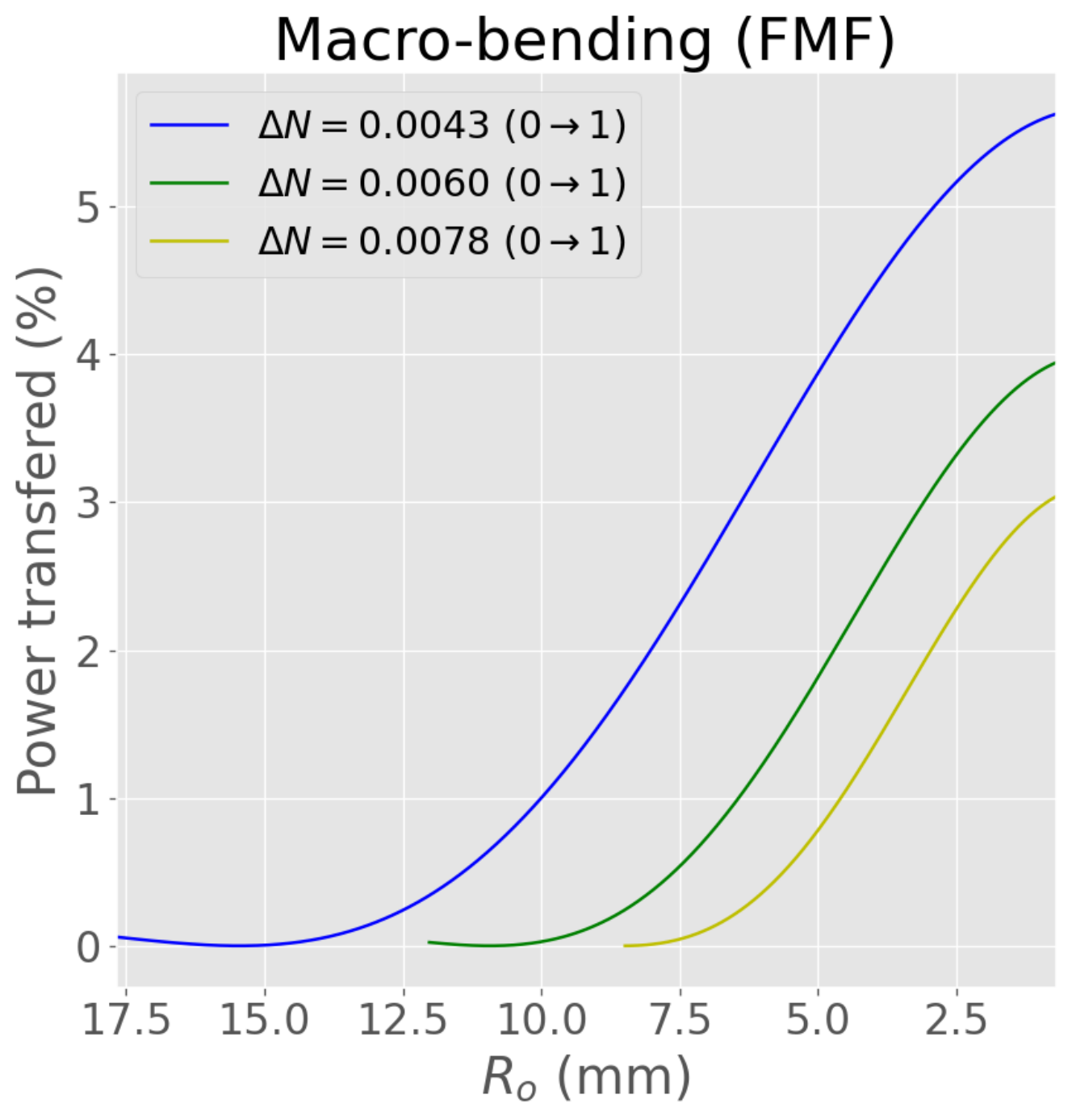}
         \caption{ }
         \label{figtap-b}
     \end{subfigure}
\caption{(a) Image of a macro-bending for tapping a multimode optical fibre.
 (b) Power transfered  between modes $m=0$ and $m=1$  by macro-bending tapping as a function of curvature radius $R_{o}$.}
\label{figtap}
\end{figure}

Let us consider a macro-bending of the optical fibre, then the refractive index profile is perturbed and the following perturbation $\Delta n^2$  can be used \cite{Schermer07}, 
\begin{equation}
\Delta n^{2}(x,y,z)=\frac{2(1+\kappa)x}{R(z)}n^{2}(x,y)
\end{equation}
where $\kappa$ is a coefficient dependent of the elasto-optic properties of the optical fibre [5] and $R(z)$ characterize the local curvature radius in the perturbed region by tapping as shown in Fig.\ref{figtap-a}.  We assume the following trial function describing the $z$-dependence of the local  radius $R(z)= R_{o}e^{z^{4}/d^{4}}$, where  the parameter $R_{o}$ is the bending radio at $z=0$ and $d$ is related to the reaching of the perturbation. Besides, a linear dependence between these parameters is assumed, that is, $R_{o}=a_{o}d$, where $a_{o}$ is a constant related to the bending angle of the optical fibre.  Finally, the symmetry of the perturbation allows only even-odd coupling, for example, between the fundamental and first excited modes with effective indices $N_{0}$ and $N_{1}$, and therefore a difference $\Delta N =N_{1}-N_{0}$.  Next, by using the perturbative solution given by Eq.\ref{pert-coupling} for modal coupling \cite{alex2025},   then  transferred  power curves  to the first mode, that is, $P_{1}=\vert a_{1}\vert^{2}$, as a function of bending radius $R_{o}$ can be obtained.  In Fig.\ref{figtap-b} we present some of these curves  for different  $\Delta N$.  Note that a  compromise between $\Delta N$ and transferred power is  required, that is, $\Delta N$ must be  large enough to avoid  spurious modal coupling but  small enough to obtain modal coupling by tapping. Anyway, curves show that a significant and measurable  amount of energy can be transferred to the first mode with usual values of the bending radius  to achieve tapping \textcolor{black}{(5--10\ mm)}. In short, in lines A$_{i}$-A'$_{i}$ or B'$_{i}$-B$_{i}$ we can detect the presence of an eavesdropper by monitoring the transferred power $P_{1}$.

\subsubsection{Active APLS based on multicore modal reciprocity}
Finally we present a more complex active APLS. Let us consider a MCF with $N$ cores coupled, that is, $N$ spatial modes and polarization modes H and V (2N-dimensional space), which connects points A$_{1}$ and A'$_{1}$ shown in Fig.\ref{Figinicial}. As commented, the submarine optical line A$_{1}$-A'$_{1}$ can reach points  beyond continental shelf. The MCF provides 2N modes H1, V1, H2, V2,…, HN, VN and then  we can define input and output vectors at A$_{1}$ and A'$_{1}$, that is, $\boldsymbol{V}_{A_{1}}$, $\boldsymbol{V}_{A'_{1}}$ and $\boldsymbol{U}_{A_{1}1}$, $\boldsymbol{U}_{A'_{1}}$ with 2N components corresponding to complex amplitudes of the optical field in each mode. On the other hand, it can be proven that the channel from A$_{1}$ to A'$_{1}$ can be characterized by  a matrix $H_{A_{1} A'_{1}}$ of dimension $2N{\rm x}2N$, including polarization. Such a matrix fulfils the following reciprocity property (analogue to wireless communications \cite{Ren11,Alda2020} but including polarization modal coupling) with  the matrix propagation $H_{A'_{1}A_{1}}$ from A'$_{1}$ to A$_{1}$ \cite{PRA21}
\begin{equation}
	H_{A'_{1}A_{1}}=D(H_{A_{1}1A'_{1}1}^{t})D,
\end{equation}
with $t$ indicating transpose and $D=I_{N}\otimes Z$, where $\otimes$  is the tensor product, $Z$ is the third Pauli’s matrix and $I_{N}$ the matrix identity. In the simpler case A$_{1}$ and A'$_{1}$ transmit to each other the $i$-th and $j$-th element of the input vectors $\boldsymbol{V}_{A1}$, $\boldsymbol{V}_{A'1}$  and detect the same i-th and j-th element of the output vector $\boldsymbol{U}_{A1}$, $\boldsymbol{U}_{A'1}$.  Importantly, the reciprocity property shows that A$_{1}$ and A'$_{1}$ measure the same intensity. Since these coefficients change randomly we can generate a random sequence of signals $S$. At each point A$_{1}$ and A'$_{1}$ has to be   a transmitter and a detector of a classical signal and some processing electronic devices  for classical low speed fibre communications. We can codify the intensity of signals $S$ with two levels corresponding to bits $\{0,1\}$ and therefore a classical key can be shared.  Obviously, the use of this APLS in the other lines   A$_{i}$-A'$_{i}$  will provide a larger security.

We present as an example and in an explicit  the case of two spatial modes 1 and 2, along with polarization modes $H$ and $V$, that is, four modes: $H1,V1,H2,V2$. By taking into account the reciprocity property the propagation matrices $M=H_{A_{1} A'_{1}}$ and $M'=H_{A'_{1}A_{1}}$ are given by the following expressions:
\begin{equation}
\hspace{-0,5cm} M=\begin{pmatrix}
a_{H1H1} & a_{H1V1} & a_{H1H2} & a_{H1V2} \\
a_{V1H1} & a_{V1V1} & a_{V1H2} & a_{V1V2} \\
a_{H2H1} & a_{H2V1} & a_{H2H2} & a_{H2V2} \\
a_{V2H1} & a_{V2V1} & a_{V2H2} & a_{V2V2} \\
\end{pmatrix},\,
\end{equation}
\begin{equation}
M'=\begin{pmatrix}
a_{H1H1} & -a_{V1H1} & a_{H2H1} & -a_{V2H1} \\
-a_{H1V1} & a_{V1V1} & -a_{H2V1} & a_{V2V1} \\
a_{H1H2} & -a_{V1H2} & a_{H2H2} & -a_{V2H2} \\
-a_{H1V2} & a_{V1V2} & -a_{H2V2} & a_{V2V2} \\
\end{pmatrix}.
\end{equation}
Now, let us consider that  A$_{1}$, for example,  emits the vector $\boldsymbol{V}_{A_{1}}=(0,1,0,0)^{t}$  and detects in position two, that is, projects on the same vector (0,1,0,0), and A'$_{1}$, for example,  emits the vector $\boldsymbol{V}_{A'_{1}}=(0,0,1,0)^{t}$ and detects in position three, that is, projects on the same vector $\boldsymbol{V}_{A'_{1}}$, then the following results are obtained:   the state sent by  A$_{1}$ is transformed under propagation, that is, $\boldsymbol{U}_{A_{1}}=M\boldsymbol{V}_{A_{1}}$, and  therefore at A'$_{1}$ the state becomes  $\boldsymbol{U}_{A_{1}}=(a_{H1V1}, a_{V1V1},a_{H2V1},a_{V2V1})^{t}$, and therefore A'$_{1}$   detects $\boldsymbol{U}_{A_{1}}\cdot \boldsymbol{V}_{A'_{1}}=a_{H2V1}$; on the other hand, the state sent by  A'$_{1}$ is transformed under propagation, that is,  $\boldsymbol{U}_{A'_{1}}=M'\boldsymbol{V}_{A'_{1}}$ and  then at A$_{1}$   the state becomes  $\boldsymbol{U}_{A'_{1}}=(a_{H2H1}, -a_{H2V1},a_{H2H2},-a_{H2V2})^{t}$, and therefore A$_{1}$   detects $\boldsymbol{U}_{A'_{1}}\cdot \boldsymbol{V}_{A_{1}}=-a_{H2V1}$. In short,  the same power $\vert a_{H2V1}\vert^{2}$ is detected at A$_{1}$ and A'$_{1}$, and accordingly  a classical key can be obtained. This key can be used to transmit the keys sent from  A$_{1}$. Obviously, the same APLS can be used in some others lines A$_{1}$-A'$_{1}$ and/or B$_{1}$-B'$_{1}$.

\subsubsection{Mixture of different APLS and other APLSs}
It is interesting to comment that mixtures of different APLS can be used. For example, we can combine modal power transference and SMF fibres. For example, let us consider   a SMF and a  two-core optical fibre, with a single-mode core and a few-mode core.  For the case $N=2$ one bit can be sent through the  SMF  and the second bit can be sent through the single mode core, therefore Eve has to tap two optical fibres and moreover the few-mode core can undergo modal coupling and therefore the tapping can be detected. Finally, there are more complex APLS, as for example sensing techniques such as DAS which are being proposed to potential damages or attacks to the optical lines \cite{Chen2022}, however these sensing techniques require more studies and probes.

\section{Summary}
We have presented a hybrid QKD approach based on  multichannel topologies for submarine optical communications.  \textcolor{black}{It is analogous to some approaches used  in Earth-Satellite QKD systems.} Several quantum lines on the continental shelves of Alice and Bob are installed where an autocompensating high-dimensional DM-CV-QKD protocol based on product states has been proposed. High dimensionality  provides a high secret key rate, and  the autocompensating technique reduces part of the hardware required to implement  submarine QKD, such as coherent sources, photodetectors, and so on.   Next, a local and/or global combination of secret keys in non quantum optical channels, as far away from shore as possible, provide layers of security that require Eve to perform simultaneous attacks on several channels in deep waters to break the security. \textcolor{black}{In this regard, a narrow continental shelf favours security measures and could be a factor in coastal node placement.}  Some possible topologies and keys processing methods  have been presented according to the required level of security or the technological possibilities of the existing submarine optical lines. Finally, additional physical layers of security,  both passive and active ones,  have been proposed for the  buried segment of submarine optical communications lines, which would require  Eve to perform optical tapping in very deep waters. In short, while the distances between Alice (A)  and Bob (B) for submarine optical communications  are  beyond the current QKD possibilities,  these hybrid QKD approach can be  very practical and useful solution.

\ \\

\ \\

%\begin{backmatter}
{\bf Funding}
This work was supported in part by the MICIN, European Union NextGenerationEU under Grant PRTR-C17.I1, and in part by the Galician Regional Government through Planes Complementarios de I+D+I con las Comunidades Aut\'onomas in Quantum Communication. It was also funded in part by MCIU/ AEI / 10.13039/501100011033 / FEDER, UE under project PID2023-152607NB-I00.

\ \\
%Content in the funding section will be generated entirely from details submitted to Prism. Authors may add placeholder text in this section to assess length, but any text added to this section will be replaced during production and will display official funder names along with any grant numbers provided. If additional details about a funder are required, they may be added to the Acknowledgment, even if this duplicates some information in the funding section. For preprint submissions, please include funder names and grant numbers in the manuscript.

%\bmsection{Acknowledgment}

%A. Vázquez thanks   the MICIN, European Union NextGenerationEU under Grant PRTR-C17.I1, and  the Galician Regional Government through Planes Complementarios de I + D + I con las Comunidades Autónomas in quantum communication.

%Additional information crediting individuals who contributed to the work being reported, clarifying who received funding from a particular source, or other information that does not fit the criteria for the funding block may also be included; for example, ``K. Flockhart thanks the National Science Foundation for help identifying collaborators for this work.'' 

{\bf Disclosures}
EFM: NEC Corporation (E). The remaining authors declare no conflicts of interest.

\ \\

%%%%%%%%%%%%%%%%%%%%%%% References %%%%%%%%%%%%%%%%%%%%%%%%%

%Add references with BibTeX or manually 
%\cite{Zhang:14,OPTICA,FORSTER2007,Dean2006,testthesis,Yelin:03,Masajada:13,codeexample}.

%%%%%%%%%% If using BibTeX:
\bibliographystyle{unsrt}
\bibliography{bibliog2}

@Article{Bedington2017,
author={Bedington, Robert
and Arrazola, Juan Miguel
and Ling, Alexander},
title={Progress in satellite quantum key distribution},
journal={npj Quantum Information},
year={2017},
month={Aug},
day={09},
volume={3},
number={1},
pages={30},
issn={2056-6387},
doi={10.1038/s41534-017-0031-5},
url={https://doi.org/10.1038/s41534-017-0031-5}
}

@article{Adnan2024_CV-QKD_100km,
	author = {Adnan A. E. Hajomer  and Ivan Derkach  and Nitin Jain  and Hou-Man Chin  and Ulrik L. Andersen  and Tobias Gehring },
	title = {Long-distance continuous-variable quantum key distribution over 100-km fiber with local local oscillator},
	journal = {Science Advances},
	volume = {10},
	number = {1},
	pages = {eadi9474},
	year = {2024},
	doi = {10.1126/sciadv.adi9474},
	URL = {https://www.science.org/doi/abs/10.1126/sciadv.adi9474},
	eprint = {https://www.science.org/doi/pdf/10.1126/sciadv.adi9474}}

@INPROCEEDINGS{Bian2025_40-km_Mbps_DM-CV-QKD, 
	
	author={Bian, Yiming and Fan, Lu and Xu, Xuesong and Zhao, Liang and Wu, Mingze and Yu, Song and Zhang, Yichen},
	
	booktitle={2025 Optical Fiber Communications Conference and Exhibition (OFC)}, 
	
	title={40-km Mbps Discrete-Modulated Continuous Variable Quantum Key Distribution With Constellation Shaping Pre-Optimization}, 
	
	year={2025},
	
	volume={},
	
	number={},
	
	pages={1-3},
	
	doi={}}

@book{chesnoy2025undersea,
	title={Undersea Fiber Communication Systems},
	author={Chesnoy, J. and Antona, J.C.},
	isbn={9780443298868},
	url={https://books.google.es/books?id=nuLk0AEACAAJ},
	year={2025},
	publisher={Elsevier Science},
	edition=3
}

@article{Liao2018,
  title = {Satellite-Relayed Intercontinental Quantum Network},
  author = {Liao, Sheng-Kai and Cai, Wen-Qi and Handsteiner, Johannes and Liu, Bo and Yin, Juan and Zhang, Liang and Rauch, Dominik and Fink, Matthias and Ren, Ji-Gang and Liu, Wei-Yue and Li, Yang and Shen, Qi and Cao, Yuan and Li, Feng-Zhi and Wang, Jian-Feng and Huang, Yong-Mei and Deng, Lei and Xi, Tao and Ma, Lu and Hu, Tai and Li, Li and Liu, Nai-Le and Koidl, Franz and Wang, Peiyuan and Chen, Yu-Ao and Wang, Xiang-Bin and Steindorfer, Michael and Kirchner, Georg and Lu, Chao-Yang and Shu, Rong and Ursin, Rupert and Scheidl, Thomas and Peng, Cheng-Zhi and Wang, Jian-Yu and Zeilinger, Anton and Pan, Jian-Wei},
  journal = {Phys. Rev. Lett.},
  volume = {120},
  issue = {3},
  pages = {030501},
  numpages = {4},
  year = {2018},
  month = {Jan},
  publisher = {American Physical Society},
  doi = {10.1103/PhysRevLett.120.030501},
  url = {https://link.aps.org/doi/10.1103/PhysRevLett.120.030501}
}

@article{Salvail2010,
author = {Louis Salvail and Momtchil Peev and Eleni Diamanti and Romain Alléaume and Norbert Lütkenhaus and Thomas Länger},
title ={Security of trusted repeater quantum key distribution networks},
journal = {Journal of Computer Security},
volume = {18},
number = {1},
pages = {61-87},
year = {2010},
doi = {10.3233/JCS-2010-0373},
URL = { https://journals.sagepub.com/doi/abs/10.3233/JCS-2010-0373},
eprint = { https://journals.sagepub.com/doi/pdf/10.3233/JCS-2010-0373}
}

@inproceedings{Suboptic1,
  author={Mateo, E.F. and   Liñares, J. and  Prieto-Blanco, X. and  Inada, Y},
  booktitle={Suboptic 2023, Conference Proceedings}, 
  title={A hybrid quantum cryptography method for submarine optical communications}, 
  year={2023},
  volume={},
  number={},
  pages={ WE3A:1-5},
  doi={}}

@article{Sheridan2010,
  title = {Security proof for quantum key distribution using qudit systems},
  author = {Sheridan, Lana and Scarani, Valerio},
  journal = {Phys. Rev. A},
  volume = {82},
  issue = {3},
  pages = {030301},
  numpages = {4},
  year = {2010},
  month = {Sep},
  publisher = {American Physical Society},
  doi = {10.1103/PhysRevA.82.030301},
  url = {https://link.aps.org/doi/10.1103/PhysRevA.82.030301}
}

@book{schleich01,
  address = {Berlin},
  author = {Schleich, Wolfgang P.},
  publisher = {Wiley-VCH},
  title = {Quantum Optics in Phase Space},
  year = 2001
}

@book{Loudon1983,
  address = {Oxford},
  author = {Loudon, R.},
  edition = {Second},
  publisher = {Clarendon Press},
  title = {The Quantum Theory of Light},
  year = 1983
}

@book{Lee,
	Author = {Lee, D.L.},
	Title = {Electromagnetic Principles of Integrated Optics},
	Publisher = {Wiley, New York},
	Year = {1986}}

@ARTICLE{Alda2020,

  author={Aldaghri, Nasser and Mahdavifar, Hessam},

  journal={IEEE Transactions on Information Forensics and Security}, 

  title={Physical Layer Secret Key Generation in Static Environments}, 

  year={2020},

  volume={15},

  number={},

  pages={2692-2705},

  doi={10.1109/TIFS.2020.2974621}}

@ARTICLE{Chen2022,
  title    = "On-line status monitoring and surrounding environment perception
              of an underwater cable based on the phase-locked {$\Phi$-OTDR}
              sensing system",
  author   = "Chen, Xiaohong and Zou, Ningmu and Wan, Yiming and Ding, Zhewen
              and Zhang, Chi and Tong, Shuai and Lu, Yanqing and Wang, Feng and
              Xiong, Fei and Zhang, Yixin and Zhang, Xuping",
  journal  = "Opt Express",
  volume   =  30,
  number   =  17,
  pages    = "30312--30330",
  month    =  aug,
  year     =  2022,
  address  = "United States",
  language = "en"
}

@Article{Huttner2022,
author={Huttner, Bruno
and All{\'e}aume, Romain
and Diamanti, Eleni
and Fr{\"o}wis, Florian
and Grangier, Philippe
and H{\"u}bel, Hannes
and Martin, Vicente
and Poppe, Andreas
and Slater, Joshua A.
and Spiller, Tim
and Tittel, Wolfgang
and Tranier, Benoit
and Wonfor, Adrian
and Zbinden, Hugo},
title={Long-range {QKD}  without trusted nodes is not possible with current technology},
journal={npj Quantum Information},
year={2022},
month={Sep},
day={09},
volume={8},
number={1},
pages={108},
issn={2056-6387},
doi={10.1038/s41534-022-00613-4},
url={https://doi.org/10.1038/s41534-022-00613-4}
}

@inproceedings{alex2025,
  author={Vázquez, A.  and Prieto-Blanco, X. and Li{\~{n}}ares, J. and  Kurahashi, R. and  Mateo, E.F.},
  booktitle={Suboptic 2025, Conference Proceedings}, 
  title={Comparison between few-mode and multi-core fibers for protection against tapping}, 
  year={2025},
  volume={},
  number={},
  pages={NG02:1-5},
  doi={}}

@ARTICLE{Ren11,

  author={Ren, Kui and Su, Hai and Wang, Qian},

  journal={IEEE Wireless Communications}, 

  title={Secret key generation exploiting channel characteristics in wireless communications}, 

  year={2011},

  volume={18},

  number={4},

  pages={6-12},

  doi={10.1109/MWC.2011.5999759}}

@ARTICLE{Schermer07,

  author={Schermer, Ross T. and Cole, James H.},

  journal={IEEE Journal of Quantum Electronics}, 

  title={Improved Bend Loss Formula Verified for Optical Fiber by Simulation and Experiment}, 

  year={2007},

  volume={43},

  number={10},

  pages={899-909},

  doi={10.1109/JQE.2007.903364}}

@article{Bai2012,
author = {Neng Bai and Ezra Ip and Yue-Kai Huang and Eduardo Mateo and Fatih Yaman and Ming-Jun Li and Scott Bickham and Sergey Ten and Jes\'us  Li$\tilde{\rm n}$ares and and Carlos Montero and Vicente Moreno and Xes\'{u}s Prieto and Vincent Tse and Kit Man Chung and Alan Pak Tao Lau and Hwa-Yaw Tam and Chao Lu and Yanhua Luo and Gang-Ding Peng and Guifang Li and Ting Wang},
journal = {Opt. Express},
number = {3},
pages = {2668--2680},
publisher = {OSA},
title = {Mode-division multiplexed transmission with inline few-mode fiber amplifier},
volume = {20},
month = {Jan},
year = {2012},
url = {http://www.opticsexpress.org/abstract.cfm?URI=oe-20-3-2668},
}

@article{Dynes2016,
author = {J. F. Dynes and S. J. Kindness and S. W.-B. Tam and A. Plews and A. W. Sharpe and M. Lucamarini and B. Fr\"{o}hlich and Z. L. Yuan and R. V. Penty and A. J. Shields},
journal = {Opt. Express},
number = {8},
pages = {8081--8087},
publisher = {OSA},
title = {Quantum key distribution over multicore fiber},
volume = {24},
month = {Apr},
year = {2016},
url = {http://www.opticsexpress.org/abstract.cfm?URI=oe-24-8-8081},
}

@article{Bethune02,
doi = {10.1088/1367-2630/4/1/342},
url = {https://dx.doi.org/10.1088/1367-2630/4/1/342},
year = {2002},
month = {jul},
publisher = {},
volume = {4},
number = {1},
pages = {42},
author = {Donald S Bethune and William P Risk},
title = {Autocompensating
quantum cryptography},
journal = {New Journal of Physics}}

@article{PRA21,
  title = {Fully autocompensating high-dimensional quantum cryptography by quantum degenerate four-wave mixing},
  author = {Li\~nares, Jes\'us and Prieto-Blanco, Xes\'us and Balado, Daniel and Carral, Gabriel M.},
  journal = {Phys. Rev. A},
  volume = {103},
  issue = {4},
  pages = {043710},
  numpages = {12},
  year = {2021},
  month = {Apr},
  publisher = {American Physical Society},
  doi = {10.1103/PhysRevA.103.043710},
  url = {https://link.aps.org/doi/10.1103/PhysRevA.103.043710}
}

@article{optik2025,
  title = {$2^{N}$-dimensional autocompensating discrete modulation $\text{CV-QKD}$ protocol  in optical fibers},
  author = {A. Vázquez and X. Prieto-Blanco and E.F. Mateo and J. Liñares},
  journal = { Optik (Submitted)},
  volume = {},
  issue = {},
  pages = {},
  numpages = {},
  year = {2025},
  month = {Apr},
  publisher = {},
 doi = {10.1103/PhysRevA.103.043710},
 url = {https://link.aps.org/doi/10.1103/PhysRevA.103.043710}
}

@article{Namiki2003,  
author={ R. Namiki and T. Hirano},  
journal={Phys. Rev. A},   
title={Security of quantum cryptography using balanced homodyne detection},   
year={2003},  
volume={67},
pages={022308-1-7)}}

@article{Pirandola2020,
author = {S. Pirandola and U. L. Andersen and L. Banchi and M. Berta and D. Bunandar and R. Colbeck and D. Englund and T. Gehring and C. Lupo and C. Ottaviani and J. L. Pereira and M. Razavi and J. Shamsul Shaari and M. Tomamichel and V. C. Usenko and G. Vallone and P. Villoresi and P. Wallden},
journal = {Adv. Opt. Photon.},
number = {4},
pages = {1012--1236},
publisher = {Optica Publishing Group},
title = {Advances in quantum cryptography},
volume = {12},
month = {Dec},
year = {2020},
url = {https://opg.optica.org/aop/abstract.cfm?URI=aop-12-4-1012},
doi = {10.1364/AOP.361502},
}

@INPROCEEDINGS{Kawamoto2005,
  author={Kawamoto, Y. and Hirano, T. and Namiki, R. and Ashikaga, M. and Shimoguchi, A. and Ohta, K.},
  booktitle={International Quantum Electronics Conference, 2005.}, 
  title={Plug and play systems for quantum cryptography with continuous variables}, 
  year={2005},
  volume={},
  number={},
  pages={1612-1614},
  doi={10.1109/IQEC.2005.1561132}}

@article{PLUG97,
author = {Muller,A.  and Herzog,T.  and Huttner,B.  and Tittel,W.  and Zbinden,H.  and Gisin,N. },
title = {Plug and play systems for quantum cryptography},
journal = {Applied Physics Letters},
volume = {70},
number = {7},
pages = {793-795},
year = {1997},
doi = {10.1063/1.118224},

URL = { 
        https://doi.org/10.1063/1.118224
    
},
eprint = { 
        https://doi.org/10.1063/1.118224}
}

@article{Bal19,
author = {Daniel Balado and Jes\'{u}s Li{\~{n}}ares and Xes\'{u}s Prieto-Blanco and David Barral},
journal = {J. Opt. Soc. Am. B},
number = {10},
pages = {2793--2803},
publisher = {Optica Publishing Group},
title = {Phase and polarization autocompensating {$N$}-dimensional quantum cryptography in multicore optical fibers},
volume = {36},
month = {Oct},
year = {2019},
url = {https://opg.optica.org/josab/abstract.cfm?URI=josab-36-10-2793},
doi = {10.1364/JOSAB.36.002793},
}

%%%%%%%%%% If preparing manually:
% \begin{thebibliography}{1}
% \newcommand{\enquote}[1]{``#1''}

% \bibitem{Zhang:14}
% Y.~Zhang, S.~Qiao, L.~Sun, Q.~W. Shi, W.~Huang, L.~Li, and Z.~Yang,
%   \enquote{Photoinduced active terahertz metamaterials with nanostructured
%   vanadium dioxide film deposited by sol-gel method,}
%   {\protect\JournalTitle{Optics Express}} \textbf{22}, 11070--11078 (2014).

% \bibitem{Optica}
% {Optica}, \enquote{{Optica Publishing Group},}
%   \url{http://www.opg.optica.org}.

% \bibitem{FORSTER2007}
% P.~Forster, V.~Ramaswamy, P.~Artaxo, T.~Bernsten, R.~Betts, D.~Fahey,
%   J.~Haywood, J.~Lean, D.~Lowe, G.~Myhre, J.~Nganga, R.~Prinn, G.~Raga,
%   M.~Schulz, and R.~V. Dorland, \enquote{Changes in atmospheric consituents and
%   in radiative forcing,} in \enquote{Climate Change 2007: The Physical Science
%   Basis. Contribution of Working Group 1 to the Fourth Assesment Report of
%   Intergovernmental Panel on Climate Change,}  S.~Solomon, D.~Qin, M.~Manning,
%   Z.~Chen, M.~Marquis, K.~B. Averyt, M.~Tignor, and H.~L. Miler, eds.
%   (Cambridge University Press, 2007).

% \end{thebibliography}

\end{document}